\documentclass{ws-procs10x7}
 
\usepackage{balance,epsfig}

\newcolumntype{d}[1]{D{.}{.}{#1}}

\def\draftnote{}

\makeindex
\begin{document}

\title{Status of SM calculations of $b \rightarrow s $ transitions}

\author{Tobias Hurth\footnote{Heisenberg Fellow}}

\address{Department of Physics, Theory Unit, CERN, CH-1211 Geneva 23, Switzerland\\ SLAC, Stanford University, Stanford, CA 94309, USA}

\twocolumn[\maketitle\abstract{We report recent progress in SM calculations 
of $b\rightarrow s$ transitions. We discuss the first NNLL prediction of the  $\bar B \rightarrow X_s \gamma$ branching ratio, including    
important additional subtleties due to non-perturbative corrections 
and logarithmically-enhanced cut effects, and also  recent results 
on the  inclusive mode $\bar B \rightarrow X_s \ell^+ \ell^-$. 
Moreover, new results on the corresponding exclusive modes are reviewed.
Finally, we comment on the present status of the 
so-called $B \rightarrow K \pi$
puzzle in hadronic $b\rightarrow s$ transitions.}  
\keywords{Rare $B$\/-meson decays.\hfill {\tt *Heisenberg Fellow.\,\,}
{\tt CERN-PH-TH/2007-056;SLAC-PUB-12396}}]

%%%%%%%%%%%%%%%%%%%%%%%%%%%%%%%%%%%%%%%%%%%%%%%%%%%%%%%%%%%%%%%%%%%%%%%%
\section{Introduction}
In any viable new physics model we have to understand the 
important flavour problem, namely why flavour-changing neutral currents are 
suppressed.  Rare decays and CP \mbox{violating} observables exclusively 
allow an analysis of  this problem.
However, if new physics does not show up in flavour physics through 
large deviations, as recent experimental data indicate, the focus on 
theoretically clean observables within the indirect search for 
new physics  is mandatory. This also calls for more precise 
SM calculations in the first  place.

A crucial problem in the new physics search within flavour physics is 
the optimal separation of new physics effects from hadronic uncertainties.
It is well known that inclusive decay modes are dominated by partonic 
contributions; non-perturbative corrections are in general 
rather small~\cite{Hurth:2003vb}. Also ratios of exclusive  
decay  modes such as 
asymmetries are well suited for the new-physics search. Here large 
parts of the hadronic  uncertainties partially cancel out; for example, 
there are CP asymmetries  that are governed by one weak phase only; thus 
the hadronic matrix  elements cancel out completely.

Data from $K$ and $B_d$ physics show  that new sources of flavour violation
in \mbox{$s \rightarrow d$} and \mbox{$b \rightarrow d$} 
are strongly constrained, while 
the possibility of sizable  new 
contributions to \mbox{$b \rightarrow s$}  
remains open~\cite{Silvestrini:2005zb,Hurth:2003th}. We also have hints from 
model building: flavour models are not very effective in  constraining 
the \mbox{$b\rightarrow s$}  sector~\cite{Masiero:2001cc}. 
Moreover, in SUSY-GUTs 
the large mixing angle in the neutrino sector relates to  large mixing in 
the right-handed $b$-$s$ sector~\cite{Moroi:2000tk,Chang:2002mq,Harnik:2002vs}.

In the following  we discuss recent progress on  several 
\mbox{$b \rightarrow s$} observables.

\section{$\bar B \rightarrow X_s \gamma$}

Among the flavour-changing current processes, the inclusive 
$b \rightarrow s \gamma$
mode is still the most prominent. 
The stringent bounds obtained from this  mode
on various non-standard scenarios are a clear example
of the importance of clean FCNC observables in discriminating
new-physics models.
Its branching ratio  has already been measured 
by several independent experiments
using semi-inclusive or fully inclusive methods~\cite{Chen:2001fj,Abe:2001hk,Koppenburg:2004fz,Aubert:2005cu,Aubert:2006gg}. 
The world average of those five measurements (performed  by
the Heavy Flavour Averaging Group (HFAG)~\cite{unknown:2006bi}) 
for a photon energy cut $E_{\gamma} >
1.6\;{\rm GeV}$ reads
\begin{eqnarray}
{\cal B}(\bar{B} \to X_s \gamma)_{\rm exp} =&  \nonumber\\
&\hspace{-2.5cm}= \left(3.55\pm 0.24{\;}^{+0.09}_{-0.10}\pm0.03\right)\times 10^{-4}
\label{hfag}
\end{eqnarray}
where the errors are combined statistical and systematic, systematic 
due to the extrapolation, and due to the 
$b \rightarrow d\gamma$ fraction. 
\begin{figure}[t]
\begin{center}
\includegraphics[width=6.5cm,angle=0]{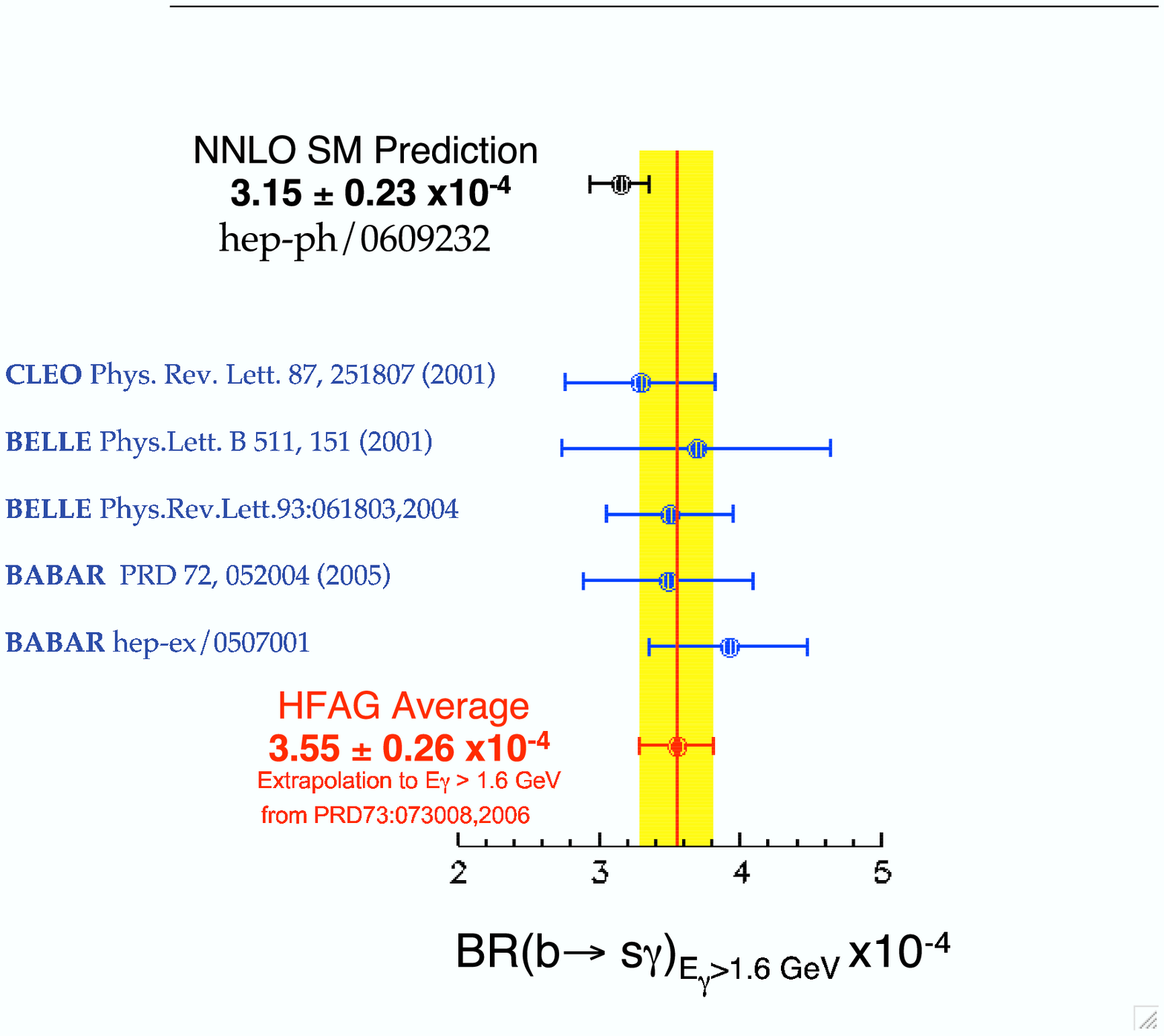}
\vspace{-0.3cm}
\caption{\sf New NNLL prediction versus HFAG average.\label{Fig:HFAG}} 
\end{center} 
\end{figure}

On the  theory side,  perturbative QCD contributions to the decay rate   
are dominant  and lead to large logarithms  
$ \alpha_s  \times$ $\log({m_b^2} / {M_W^2}) $,  which have to be resummed
in order to get a reasonable result. 
Resumming all the terms of the form 
$(\alpha_s)^p$ $(\alpha_s \log(m_b/M))^n$
(with  $M=m_t$ or $M=m_W$, $n=0,1,2,...\,\,$) for fixed $p$ corresponds for 
$p=0$ to leading-log (LL), for $p=1$ to next-to-leading-log (NLL), and  
for  $p=2$ to next-to-next-to-leading-log (NNLL) precision.  
The previous NLL prediction,  
 based on the original QCD calculations of several groups~\cite{Adel,GHW,Mikolaj,GH,Burasnew,Paolonew,Buras:2002tp,Asatrian:2004et,Ali:1990tj,Pott:1995if,Kagan:1998ym}, had an additional charm mass
renormalization scheme ambiguity, first analysed   
in Ref.\cite{Gambino:2001ew}.
 For an 
energy cut  $E_{\gamma} > 1.6\;{\rm GeV}$ it reads~\cite{Hurth:2003dk}
\begin{eqnarray}
{\cal B}(\bar{B} \to X_s \gamma)_{\rm NLL} =&  \nonumber\\
&\hspace{-3.5cm}=({3.61}   {}^{+0.24}_{-0.40}  
                     \pm 0.02  \pm 0.25  \pm 0.15 ) \times 10^{-4}
\end{eqnarray}
where the errors are due to the charm scheme dependence, CKM input,
further parametric dependences, and to perturbative scale dependence. 
The dominant  uncertainty related to the definition of $m_c$ was taken 
into account
by varying $m_c/m_b$ in the conservative range $0.18 \le m_c/m_b \le 0.31$,
which covers both, the pole mass value (with its numerical error) and 
the running mass value 
$\bar{m}_c(\mu_c)$ with $\mu_c \in [m_c,m_b]$; for the central value
$m_c/m_b = 0.23$ was used~\cite{Hurth:2003dk}. 
However, the renormalization scheme for $m_c$ is an NNLL issue. It was shown 
that a complete NNLL calculation reduces this large  uncertainty 
at least by a factor of 2~\cite{Asatrian:2005pm}.

Within a global effort,  such a NNLL calculation was quite recently 
 finalized~\cite{Misiak:2006zs}. 
The  calculational steps were performed by 
various groups~\cite{Misiak:2004ew,Gorbahn:2004my,Gorbahn:2005sa,Czakon:2006notyet,Blokland:2005uk,Melnikov:2005bx,Asatrian:2006ph,Asatrian:2006sm,Bieri:2003ue,Misiak:2006justnow}. 
One crucial piece is the calculation of the three-loop matrix elements of
the four-quark operators,  which was first made 
within the so-called large-$\beta_0$ approximation~\cite{Bieri:2003ue}. 
A calculation that goes 
beyond this
approximation by employing an interpolation in the charm quark mass $m_c$ 
from $m_c > m_b$ to the \mbox{physical} $m_c$ value has just been 
completed~\cite{Misiak:2006justnow}. It is that
part of the NNLL calculation where there is still space for improvement.

All those results lead to the first estimate of the $\bar B \rightarrow X_s \gamma$ branching ratio to NNLL precision. It reads for a photon energy cut 
$E_{\gamma} > 1.0\;{\rm GeV}$~\cite{Misiak:2006zs}:
\begin{eqnarray}\label{final1}
{\cal B}(\bar{B} \to X_s \gamma)_{\rm NNLL} =&  \nonumber\\
&\hspace{-2cm}(3.27 \pm 0.23) \times 10^{-4}.
\end{eqnarray}
%\begin{equation} \label{final1}
%{\cal B}({\bar B}\to X_s\gamma)_{\rm NNLL}= (3.27 \pm 0.23) \times 10^{-4}.
%\end{equation}
The overall uncertainty consists of 
non-perturbative (5\%), parametric (3\%), higher-order
(3\%) and $m_c$-interpolation ambiguity (3\%), which have been added 
in quadrature.
For higher  photon energy cut we have the following 
numerical fit: 
\begin{equation}
\left( \frac{{\cal B}(E_\gamma > E_0)}{
             {\cal B}(E_\gamma > 1.0\;{\rm GeV})} \right)
\simeq 1-0.031y-0.047y^2,
\end{equation}
where $y = E_0/(1.0\,{\rm GeV}) -1$. This formula coincides with the NNLL 
results up to $\pm 0.1\%$ for $E_0 \in [1.0,\;1.6]\;$GeV. The error is
practically $E_0$-independent in this range. For 
$E_{\gamma} > 1.6\;{\rm GeV}$ the NNLL prediction reads~\cite{Misiak:2006zs}:
\begin{eqnarray}\label{final2}
{\cal B}(\bar{B} \to X_s \gamma)_{\rm NNLL} =&  \nonumber\\
&\hspace{-2cm} (3.15  \pm 0.23) \times 10^{-4}.
\end{eqnarray}
%\begin{equation} \label{final2}
%{\cal B}({\bar B}\to X_s\gamma)_{\rm NNLL}= (3.15  \pm 0.23) \times 10^{-4}.
%\end{equation}
Compared with the HFAG average, given in Eq.(\ref{hfag}), the NNLL prediction is $1.2 \sigma$ 
below the experimental data (see Fig.~\ref{Fig:HFAG}~\cite{buchmueller}). 
\begin{figure}[t]
\begin{center}
\includegraphics[width=5.8cm,angle=0]{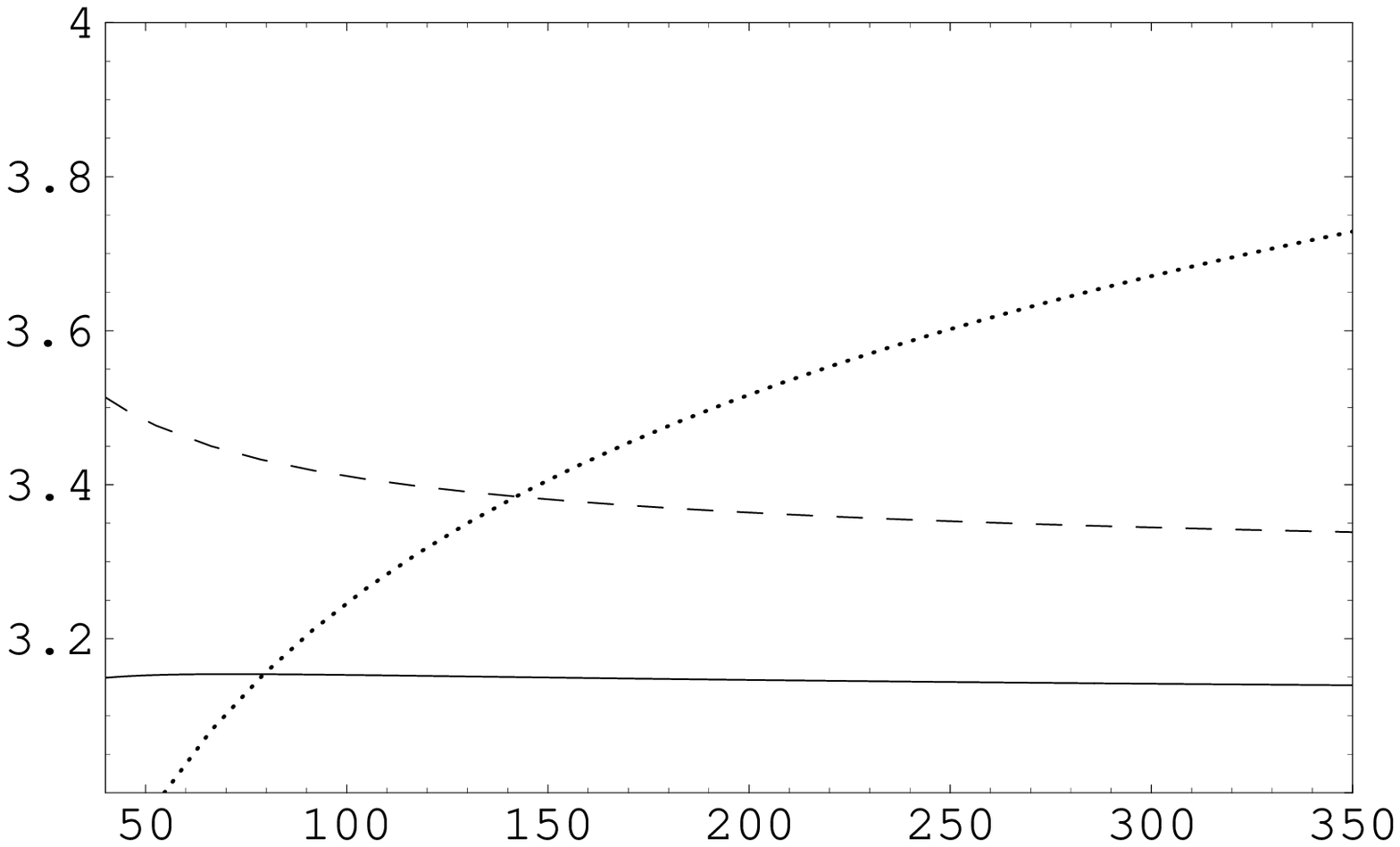}\\[2mmm]
\includegraphics[width=5.8cm,angle=0]{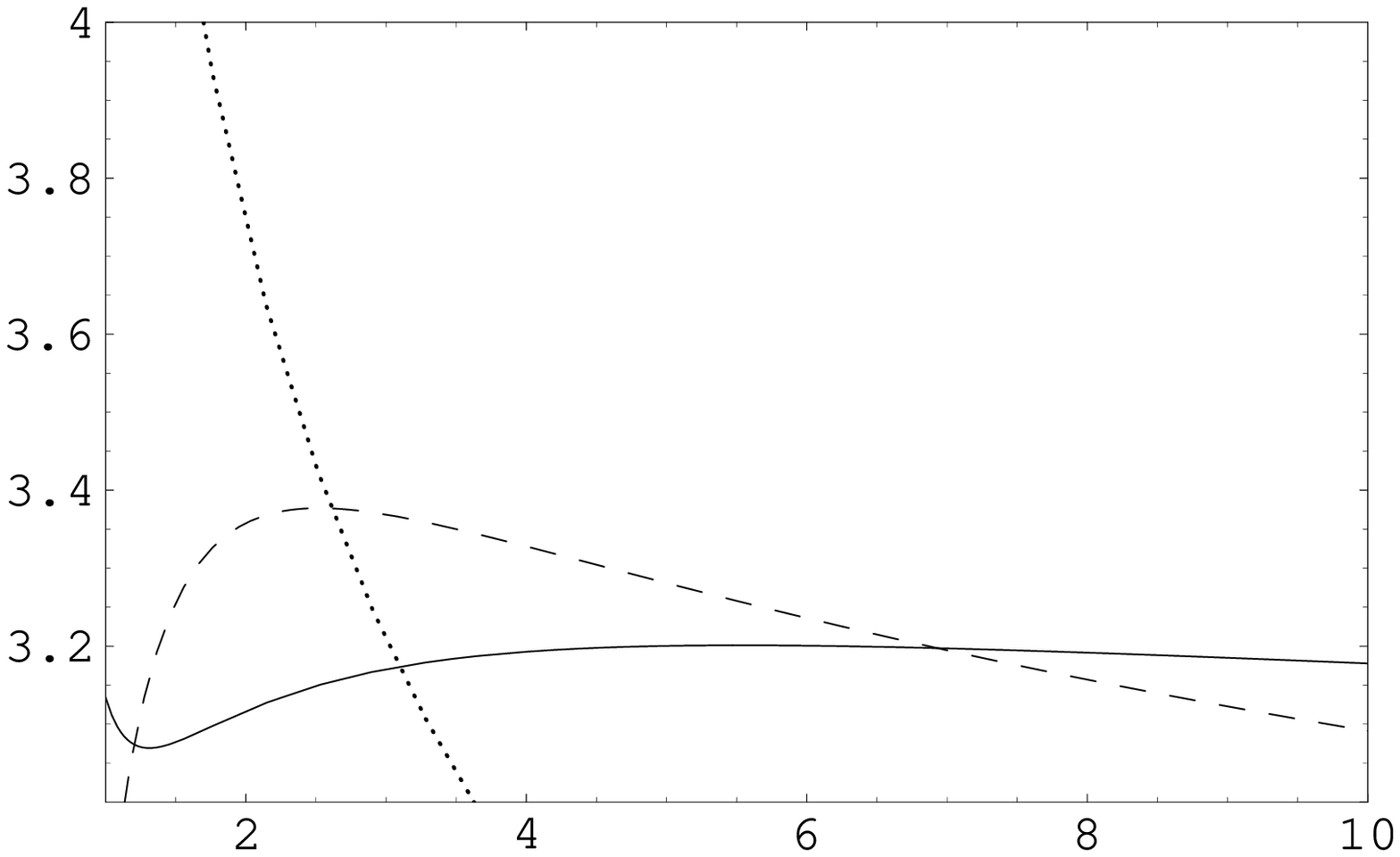}\\[2mm]
\includegraphics[width=5.8cm,angle=0]{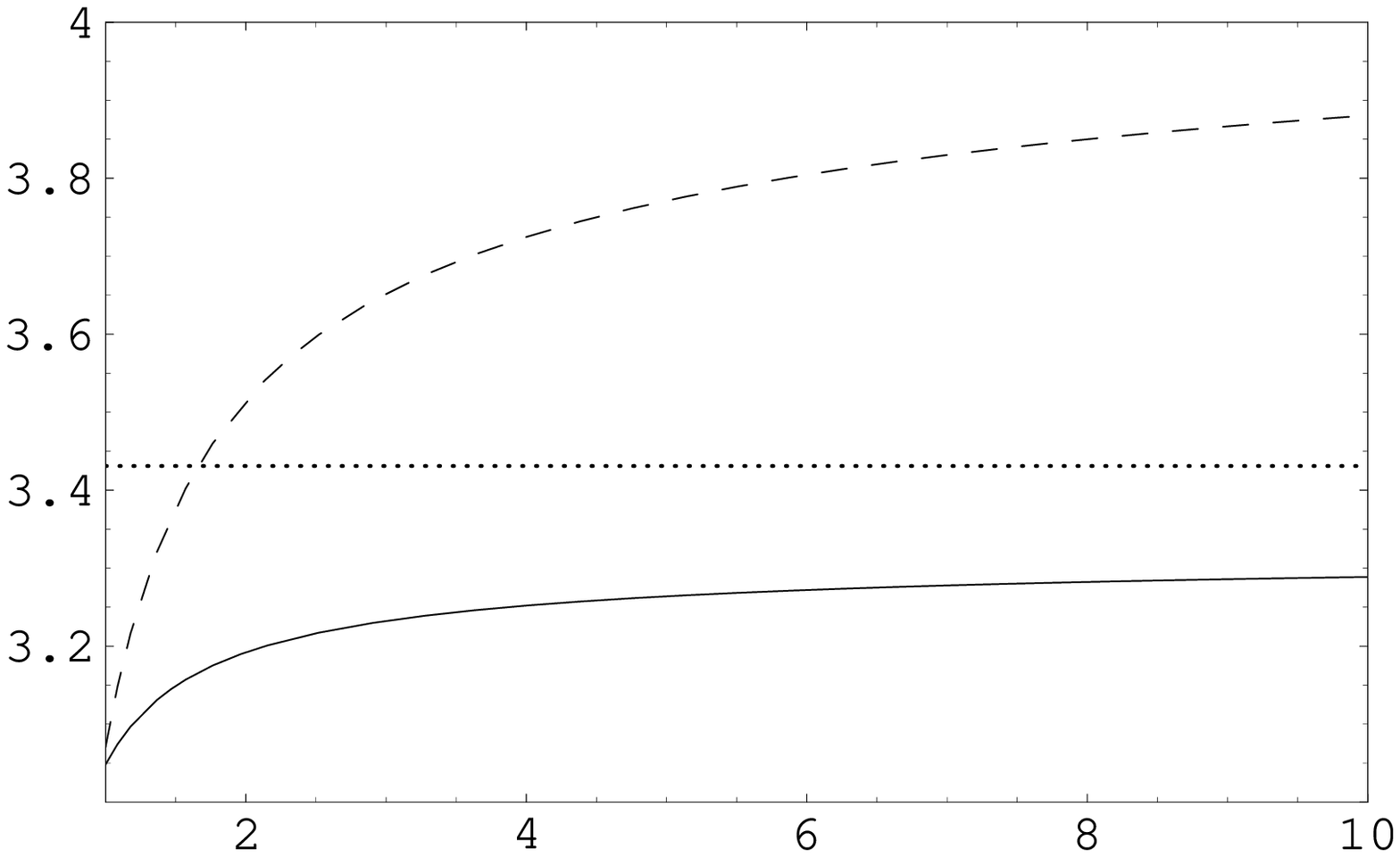}\\[-80mm]
$\mu_0\;$[GeV]\\[3.4cm]
$\mu_b\;$[GeV] \\[3.6cm]
$\mu_c\;$[GeV]\\
\caption{\sf Renormalization-scale dependence of ${\cal B}(\bar{B} \to X_s \gamma)$
  in units $10^{-4}$ at the LL (dotted lines), NLL (dashed lines) and NNLL (solid
  lines). The plots describe the dependence on the matching scale
  $\mu_0$, the low-energy scale $\mu_b$, and the charm mass renormalization
  scale $\mu_c$. \label{fig:mudep}} 
\end{center} 
\end{figure}

The reduction of the renormalization-scale dependence at the NNLL  is 
clearly seen in Fig.~\ref{fig:mudep}. 
The most important  effect occurs for 
the charm mass $\overline{\rm MS}$ renormalization scale $\mu_c$ 
that was the main source of uncertainty at the NLL. 
The current uncertainty of $\pm 3\%$ due to higher-order
(${\cal O}(\alpha_s^3)$) effects is estimated from the 
NNLL curves in Fig.~\ref{fig:mudep}. The reduction factor 
of the perturbative error is more than a factor $3$.  
The central value of the NNLL prediction is based on the 
choices $\mu_b=2.5$GeV and $\mu_c=1.5$ GeV.

There are some perturbative NNLL corrections which are not included yet 
in the present NNLL estimate, but are expected to be smaller than the 
current uncertainty: the virtual- and bremsstrahlung contributions to
the  $({\cal O}_7,{\cal O}_8)$  and $({\cal O}_8,{\cal O}_8)$ interferences at 
order $\alpha_s^2$, the NNLL bremsstrahlung contributions in the large 
$\beta_0$-approximation beyond the $({\cal O}_7,{\cal O}_7)$ interference
term (which are already available~\cite{Ligeti:1999ea}), 
the four-loop mixing of the four-quark operators into  
the operator ${\cal O}_8$ (see recent work~\cite{Czakon:2006notyet}),
the exact  $m_c$ dependence of the various matrix elements beyond the 
large $\beta_0$
approximation (see~\cite{Asatrian:2006rq}) and 
perturbative logarithmically-enhanced cut effects 
(see discussion below and \cite{Becher:2006pu}).

Nevertheless, the final result  includes subdominant contributions 
such as  the perturbative electroweak 
two-loop corrections of order $-3.7\%$~\cite{Czarnecki:1998tn,Kagan:1998ym,Baranowski:1999tq,Gambino:2000fz}
and the non-perturbative corrections scaling with  $1/m_b^2$ or $1/m_c^2$ of order 
$+1\%$ and $+3\%$ respectively~\cite{Chay:1990da,Voloshin:1996gw,Ligeti:1997tc,Grant:1997ec,Buchalla:1997ky}.

It is well known, that the local operator product
 expansion (OPE)  for the decay $\bar B \rightarrow X_s \gamma$ has certain limitations if one takes into account other operators than the  leading 
${\cal  O}_7$,  
as was already shown within the analysis of $1/m_c^2$ power corrections.     
The additional error  of $5\%$  in the NNLL prediction corresponds to   
non-perturbative corrections,  which  scale with $\alpha_s \Lambda/m_b$. 
Quite recently, a specific piece of the \mbox{additional} 
non-perturbative corrections  
was estimated~\cite{Lee:2006wn}. Because  the overall sign of the whole effect is still unknown,
this  partial estimate is not included in the central value of the 
present  NNLL prediction~\cite{Misiak:2006zs}.

\begin{figure}
\psfig{file= 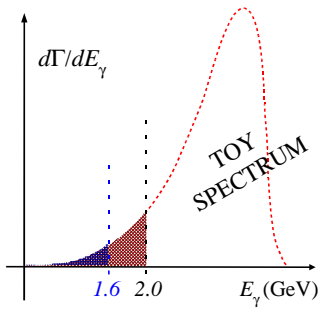,width=6.5cm} 
\vspace{-0.9cm}
\caption{\sf Cut in the photon energy spectrum}
\label{toyspectrum}
\end{figure}

However, there are more  subtleties. There is an additional sensitivity 
to non-perturbative physics, due to necessary cuts in the photon energy 
spectrum to suppress the  
background from other $B$ decays (see Fig.~\ref{toyspectrum}). 
This leads to a breakdown of the 
local OPE, which can be cured by partial resummation of these effects to 
all orders into a non-perturbative 
shape-function~\cite{Neubert:1993um,Bigi:1993ex,Mannel:1994pm}.

Those shape-function effects are taken into account in the experimental 
value by the HFAG~\cite{unknown:2006bi}
and the corresponding theoretical  uncertainties 
due to this model dependence is reflected in the extrapolation error 
in the experimental number quoted above in Eq.~(\ref{hfag}).
Here, one should keep in mind that the experimental energy cuts 
in the last  experiments are at $1.8\,{\rm GeV}$  or $1.9\,{\rm GeV}$ 
or even higher.  
The extrapolation  down to $1.6\,{\rm GeV}$ is done using three different 
theoretical schemes  to calculate the extrapolation 
factor~\cite{Benson:2004sg,Kagan:1998ym,Bosch:2004th,Buchmuller:2005zv}
and averaging those  results.

Moreover, it was argued that a cut around $1.6\,{\rm GeV}$ might not 
guarantee 
that a theoretical description in terms of a local OPE is sufficient 
because of the sensitivity to the 
scale $\Delta = m_b - 2 E_\gamma$~\cite{Neubert:2004dd}. A 
multiscale OPE with  three short-distance scales 
$m_b, \sqrt{m_b \Delta}$, and $\Delta$ 
was proposed to connect the shape function and the local OPE region. 
Quite recently,  such additional cutoff-related effects 
were  numerically estimated using (model-independent) 
SCET methods~\cite{Becher:2006pu,Becher:2005pd,Becher:2006qw}.
Those perturbative effects due to the additional scale  are negligible 
for $1.0\,{\rm GeV}$  but lead 
to an effect of order $3\%$  at $1.6\,{\rm GeV}$~\cite{Becher:2006pu}.
The size of these effects at
$1.6\,{\rm GeV}$ is at the same level as the $3\%$ higher-order uncertainty 
in the present NNLL prediction. 
It is suggestive that in the future those 
additional perturbative cut effects get analysed and combined  
together with those  already included in the experimental average 
of the HFAG.

There are also other claims for non-negligible cut effects at 
$1.6\,{\rm GeV}$~\cite{Bigi:2002qq} which, however, are based on 
models of the non-perturbative shape function.  
Moreover, there is an alternative approach to the cut  effects in the photon
energy spectrum  based on dressed gluon exponentiation and 
incorporating Sudakov and renormalon resummations~\cite{Andersen:2005bj,Andersen:2006hr}.
It should be emphasized that the higher predictive 
power of this approach is related in part to the assumption that 
non-perturbative power corrections associated with the shape function 
follow the pattern of ambiguities present in the
perturbative calculation~\cite{Gardi:2006jc}.

\section{$\bar B \rightarrow X_s \ell^+ \ell^-$}

\begin{figure}
\psfig{file=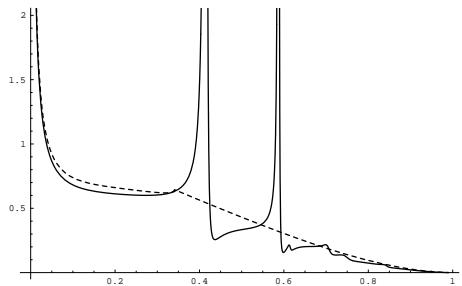,width=6cm}
\vspace{1cm} 
\caption{\sf Schematic dilepton-mass dependence of the differential branching ratio $d/ds\, BR(\bar B \rightarrow X_s \ell^+ \ell^-)$
 in units $10^{-5}$ (the dashed line corresponds to the perturbative contribution).}
\label{mikolaj}
\end{figure}

In comparison to  the  $\bar B \rightarrow X_s \gamma$, the inclusive $\bar B \rightarrow X_s \ell^+\ell^-$  decay presents a complementary and also more complex test 
of the SM.  The decay $\bar B \rightarrow X_s \ell^+\ell^-$ is particularly 
attractive because of  kinematic observables such as 
the invariant dilepton mass spectrum and the forward--backward 
(FB) asymmetry. 
These observables  are  dominated 
by perturbative  contributions if the  $c \bar c$ resonances 
that show up as large peaks in the dilepton invariant mass spectrum 
are removed by appropriate  kinematic cuts (see Fig.~\ref{mikolaj}). 
In the `perturbative $s=q^2/m_b^2$-windows', namely  
in the low-dilepton-mass  region $ 1\,{\rm GeV} < q^2 < 6\,{\rm GeV}$, and 
also in the high-dilepton-mass region with $q^2 > 14.4\,{\rm GeV}$,
theoretical predictions for the invariant mass spectrum
are dominated by the perturbative contributions;
 a theoretical precision of order $10\%$ 
is in principle possible.  
Regarding the choice of precise cuts in the dilepton mass 
spectrum, it is important that theory and  
experiment can be compared using the same energy 
cuts and any kind of
extrapolation is avoided.  

The recently calculated NNLL contributions~\cite{Asa1,Adrian2,Adrian1,Asa2,MISIAKBOBETH,Gambinonew}
have significantly 
improved the sensitivity of the inclusive $\bar B \rightarrow X_s \ell^+ \ell^-$ decay in  testing extensions of the SM in the sector of flavour 
dynamics, in particular the value of the dilepton invariant mass
$q^2_0$, for which the differential forward--backward \mbox{asymmetry}
vanishes, is one of the most precise predictions in flavour physics
with a theoretical uncertainty of order $5\%$. 

Also non-perturbative corrections scaling with $1/m_b^2$ or
$1/m_c^2$\,\,\cite{Falk,Alineu,Buchalla:1997ky,Savagenew,buchallanewnew,Bauer}
are taken into account in the present NNLL predictions.
The unknown non-perturbative corrections which scale with 
$\alpha_s\Lambda/m_b$ are less important than in the case of 
the decay $\bar B  \rightarrow X_s\gamma$. 

Recently, further refinements were presented 
such as  the NLL QED two-loop corrections to the Wilson coefficients of order 
$2\%$~\cite{Gambinonew}. Furthermore, it was shown that in the
QED two-loop corrections to matrix elements large collinear 
logarithms of the form $log(m_b/m_{\rm lepton})$ survive integration if 
only a restricted part of the dilepton mass spectrum is considered. 
This adds another contribution of order 
$+2\%$ in the low-$q^2$ region~\cite{Huber:2005ig}.

A recent update of the dilepton mass spectrum,  integrated 
over the low dilepton invariant mass region in the muonic case, 
leads to~\cite{Huber:2005ig}
\begin{equation}
{\cal B} (\bar B\to X_s \mu^+\mu^-) =   (1.59\pm 0.11)\times 10^{-6}\,,
\end{equation}
 where the error includes the parametric and perturbative uncertainties only.
For ${\cal B} (B\to X_s e^+e^-)$, in the current BaBar and Belle setups,
the logarithm of the lepton mass gets replaced by angular-cut
parameters and the integrated branching ratio for the electrons is
expected to be close to that for the muons. 
The analogous  update of the other NNLL predictions will be presented
in a forthcoming paper~\cite{future}.

There are further subtleties,  which again lead to larger theoretical 
uncertainties. In the high-${s}$ region, one encounters 
the breakdown of the 
heavy-mass expansion at the endpoint;
while the partonic contribution vanishes in the end-point,
the $1/m_b^2$ corrections tend towards a non-zero value.  
In contrast to the endpoint region of the photon energy 
spectrum in the $\bar B \rightarrow X_s \gamma$ decay,
no partial all-orders resummation into a shape function 
is possible here. However, for an integrated high-$s$ spectrum $R(s)$ an effective expansion is found in inverse powers of 
\begin{equation}
m_b^{\rm eff} = m_b \times (1 - \sqrt{s_{\rm min}})
\end{equation} rather than $m_b$. The expansion converges less
rapidly,  depending on the lower dilepton-mass cut $s_{\rm min}$~\cite{Adrian1}.

A hadronic invariant-mass cut is imposed in the present experiments 
(Babar:\, $m_X < 1.8\,{\rm GeV}$, Belle:\, $m_X < 2.0\,{\rm GeV}$)
in order to eliminate the background such as   $b \rightarrow  c\, (\rightarrow se^+\nu) e^-\bar \nu = b \rightarrow se^+e^- + \mbox{\rm missing energy}$. 
The  high-dilepton mass  region is not affected by this cut and in 
the low-dilepton mass region the kinematics with a jet-like $X_s$ and 
$m_X^2 \leq m_b \Lambda_{\rm QCD}$ implies the relevance of the 
shape function. 
A recent SCET analysis shows that using the universality of jet and 
shape functions  the $10-30\%$ reduction of the dilepton mass spectrum 
can be accurately computed using  the $\bar B \rightarrow X_s \gamma$ shape 
function. Nevertheless effects of subleading shape functions lead to
an additional uncertainty of $5\%$~\cite{Lee:2005pk,Lee:2005pw}.

It is well-known, that the measurement of the dilepton mass spectrum, 
the zero of the forward-backward asymmetry, and the $\bar B \rightarrow
X_s \gamma$ branching ratio allows to fix the magnitude and sign of all
relevant Wilson coefficients within the SM.  
In view of the fact that at present only restricted data sets are available,
it was recently proposed to focus  on  quantities which 
are integrated (over $q^2$); 
besides the total rate and the integrated forward-bachward asymmetry it
was shown that the third angular decomposition within the 
$b \rightarrow s \ell^+\ell^-$  mode is sensitive to a
different combination of Wilson coefficients~\cite{Lee:2006gs}.

\section{Exclusive $b \rightarrow s$ transitions}  

The corresponding rare exclusive decays, such as $B \to K^* \gamma$, 
$B \to K^*  \mu^+ \mu^-$ or also   
$B_s  \to \phi \gamma$, $B_s \to \phi \mu^+ \mu^-$, 
are well-accessible at the forthcoming 
LHCb experiment. 
In contrast to the measurement of the branching ratios, 
measurements of CP, forward-backward, and isospin asymmetries are 
less  sensitive to hadronic uncertainties.

For example, the value of the dilepton invariant mass in  
$B \to K^*  \mu^+ \mu^-$,
$q_0^2$, for which the differential 
forward--backward asymmetry vanishes, can be predicted in quite a clean way. 
In the QCD factorization approach, at leading order in $\Lambda_{\rm QCD}/m_b$, the value of  $q_0^2$ is free from hadronic
uncertainties at order $\alpha_s^0$, a  dependence 
on the soft form factor $\xi_\perp$ and the light-cone wave functions of 
the $B$ and $K^*$ mesons appear at order $\alpha_s^1$. The latter contribution, calculated within the  QCD factorization approach,
leads  to a large  shift (see~\cite{Beneke:2001at,Beneke:2004dp,Ali:2006ew}).
Nevertheless, there is the well-known issue of power corrections ($\Lambda_{QCD}/m_b$) 
within the QCD factorization approach which increases the theoretical uncertainty. 

An extension of the QCD factorization formula to the non-resonant decay 
$B \rightarrow K\pi \ell^+\ell^-$ with an energetic $K\pi$ pair and also
with an energetic kaon and a soft pion was presented~\cite{Grinstein:2005ud}.
Here  one relies on the fact that the forward-backward 
asymmetry is due to the interference of the helicity $J=1^+,1^-$ 
amplitudes  induced by  the 
$b \rightarrow s$ current. So it seems that no angular analysis  
is necessary to disentangle vector and tensor final states; however, 
the dependence of the non-perturbative input functions on the 
kinematic variables might differ for the two cases.
This suggests that a restriction to the resonant states only is still
the theoretically cleanest option.

There are also  certain transversity amplitudes in $B \to K^* \mu^+ \mu^-$, 
in which the hadronic formfactors also cancel out at leading order. Thus, 
such observables 
are rather insensitive to hadronic uncertainties, but  
highly sensitive to non-standard chiral  structures of the 
$b \to s$ current~\cite{Kruger:2005ep,Lunghi:2006hc}. 

Quite recently, branching ratios, isospin and CP asymmetries in exclusive  
radiative decays like $B \to K^* \gamma$ and $B_s  \to \phi \gamma$
were estimated combining QCD factorization results with
QCD sum rule estimates of power corrections, namely long-distance 
contributions due to photon and soft-gluon emisson from 
quark loops~\cite{Ball:2006eu}. 
Particulary, this leads to an estimate of the 
time-dependent CP asymmetry  in $B^0\to K^{*0}\gamma$ of                        $S=-0.022\pm 0.015^{+0}_{-0.01}$~\cite{Ball:2006eu}. 
The contribution due to soft gluon emssion is estimated 
within the QCD sum rule approach to be very small, 
$S^{sgluon} = - 0.005 \pm 0.01$, while 
a \mbox{conservative} dimensional estimate 
of this contribution due to a nonlocal  SCET operator \mbox{series} 
leads to $|S^{sgluon}| 
\approx  0.06$~\cite{Grinstein:2004uu,Grinstein:2005nu}.
Furthermore, one finds  a larger time-dependent CP asymmetry  
of around $10\%$ within the inclusive mode~\cite{Grinstein:2004uu}; 
however, the SCET estimate shows that the expansion parameter is 
$\Lambda_{QCD}/Q$ where $Q$ is the kinetic energy of the hadronic part,
while there is no contribution  at leading order. Therefore, the effect is 
expected to be larger for larger invariant hadronic mass, thus, 
the $K^*$ mode  has to have the smallest effect, below the 
`average' $10\%$~\cite{Grinstein:2004uu}.

\section{Comments on  the so-called $B \rightarrow K \pi$ puzzle}

The $B \to K \pi$ modes are well known for  being  sensitive  
to new electroweak $b \rightarrow s$ penguins beyond 
the SM~\cite{Fleischer:1997ng,Grossman:1999av}.
The data on CP-averaged $K\pi$ branching ratios can be expressed in 
terms of three ratios:
\begin{eqnarray}
{R}=\frac{\tau_{B^+}}{\tau_{B^0}}
    \frac{{\cal  B}[B^0\to\pi^-K^+]+{\cal B}[\bar B^0\to\pi^+K^-]}
        {{\cal B}[B_d^+\to\pi^+K^0]+{\cal B}[B_d^-\to\pi^-\bar K^0]}\nonumber\\
 {R_n}=\frac12 \frac{{\cal B}[B^0\to\pi^-K^+]+{\cal B}[\bar B^0\to\pi^+K^-]}
        {{\cal B}[B^0\to\pi^0K^0]+{\cal B}[\bar B^0\to\pi^0 \bar K^0]}\nonumber\\
 {R_c}=2\frac{{\cal B}[B_d^+\to\pi^0 K^+]+{\cal B}[B_d^-\to\pi^0 K^-]}
        {{\cal B}[B_d^+\to\pi^+ K^0]+{\cal B}[B_d^-\to\pi^- \bar K^0]}\nonumber
\end{eqnarray}
The actual data presented at ICHEP06 read~\cite{HFAG,Abe:2006qx,Aubert:2006ap}
\begin{eqnarray}\nonumber
R=0.92^{+0.05}_{-0.05} R_n=1.00^{+0.07}_{-0.07} R_c=1.10^{+0.07}_{-0.07}
\end{eqnarray}
\begin{figure}
\psfig{file=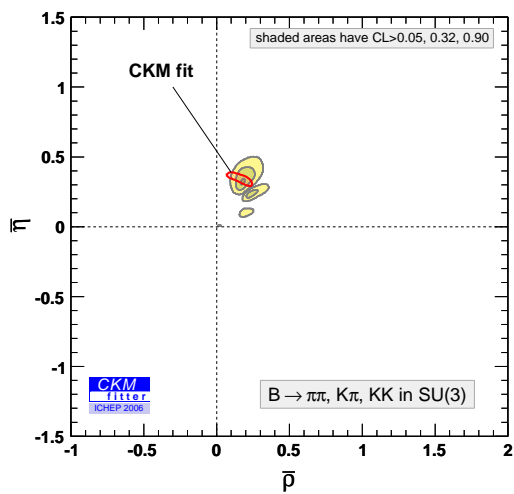,width=7cm}
\caption{\sf Constraint in the $(\bar\rho,\bar\eta)$ plane induced by the 
$\pi\pi,K\pi,K\bar K$ data compared with the standard CKM fit.}
\label{charles2}
\end{figure}

One should emphasize that in the previous analyses  
the radiative electromagnetic 
corrections to charged particles in the final state 
were  not taken into account,  as 
was emphasized in the past (see for example Ref.~\cite{Hurth:2005wg}). 
These corrections, worked out in Ref.~\cite{Baracchini:2005wp},  
are now properly included in the analysis of both 
experiments.

The present data is compatible with  the  approximate sum rule 
proposed in Refs.~\cite{Lipkin:1998ie,Gronau:1998ep,Matias:2001ch}, 
which leads to the prediction  $R_c \approx R_n$.
The data is also in agreement with
the available  SM approaches to these data based on QCD factorization 
techniques and 
on $SU(3)_F$ symmetry assumptions~\cite{Gronau:1990ka,Nir:1991cu}.
For example,  the BBNS predictions,  
based on the QCD factorization 
approach~\cite{Beneke:2003zv,BBNSprivate}, are
\begin{eqnarray} \nonumber
R=0.91^{+0.13}_{-0.11} R_n=1.16^{+0.22}_{-0.19} R_c=1.15^{+0.19}_{-0.17}
\end{eqnarray}

The latest $SU(3)_F$ analysis of 
the CKM fitter group~\cite{charlesref},
includes 
all  available $\pi\pi,K\pi,K\bar K$ modes;
also so-called  annihilation/exchange topologies and  
factorizable $SU(3)_F$ breaking are taken into account.

\begin{figure}[t]
\begin{center}
\includegraphics[width=6.4cm,height=6.8cm,angle=90]{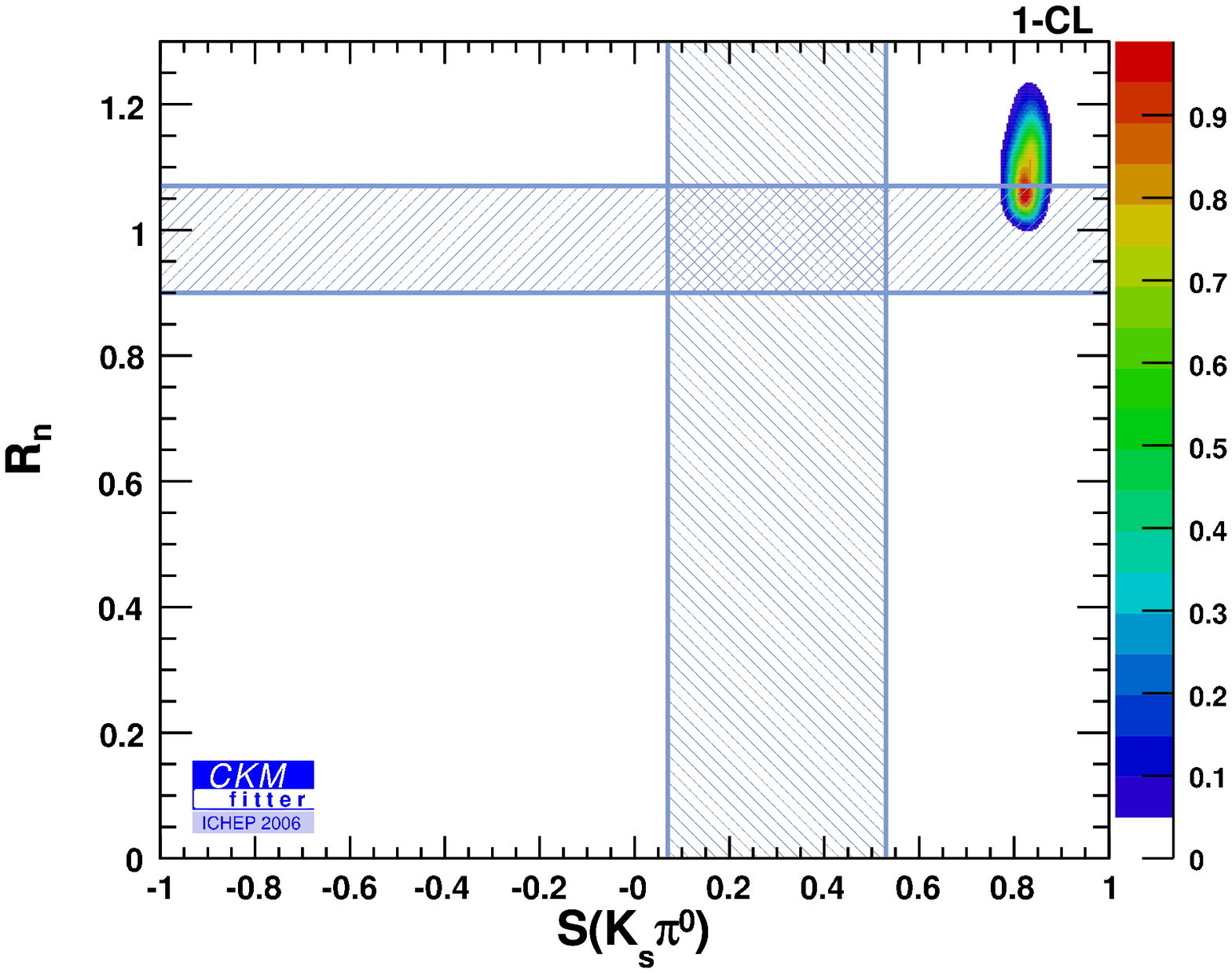}
\caption{\sf   Comparison of  direct  measurements of $R_n$ and  $S(K^0_S\pi^0)$
($1\sigma$ band) with the indirect fit ($2\sigma$ contour).   \label{charles1}} 
\end{center} 
\end{figure}

As shown  in Fig.~\ref{charles2}~\cite{charlesref}
the constraint in the $(\bar\rho,\bar\eta)$ plane induced by these data
implies that the compatibility with  the $SU(3)$ and 
SM hypothesis is very good (the so-called pValue of that 
SM analysis is of order  $30-40\%$). 
But  the $\chi^2_{\rm min}$ is not always the best measure of the 
compatibility of the data with the theory. Among the main 
contributions to the $\chi^2$  there  is 
the ratio $R_n$  
and the CP asymmetry $S(K^0_S\pi^0)$. The latter quantity 
is defined via 
\begin{eqnarray}
   a_{CP}[K^0_S\pi^0](t)&=&\\
&&\hspace{-3cm}   S(K^0_S\pi^0) {\rm sin}(\Delta m_d  t) - C(K^0_S\pi^0) {\rm cos}(\Delta m_d  t). \nonumber 
\end{eqnarray}
Both observables are very sensitive to new electroweak penguins.    
After removing them from the global fit,  
Fig.~\ref{charles1}~\cite{charlesref}  
shows the  comparison 
of the indirect fit ($2\sigma$ contour),  with $\bar \rho, \bar \eta$ 
from the CKM fit and  all  other available modes, with the
direct measurements  ($1\sigma$ band) using the new  
data. 
The indirect prediction for $R_n$ is now in good agreement with the 
direct measurement. There is still a small tension 
in the case of the observable $S(K^0_S\pi^0)$ which is at present not 
really 'puzzling' from the statistical point of view. 
Fig.~\ref{charles1}  shows that these two quantities are not 
correlated, so that  possible deviations from the SM values could 
have a completely different origin. 
These findings were just recently confirmed in a similar
approach~\cite{Fleischer:2007mq}.

Because of  the large non-factorizable contributions identified 
in the $B \to \pi\pi$ channel, large non-factorizable 
$SU(3)_F$ - or isospin-violating QCD and QED effects within the SM
cannot be ruled out at present~\cite{Feldmann:2004mg}. 
Future data from the $B$ factories and LHCb  will clarify the situation 
completely. There will be  up to 38 measured observables  depending on the 
same 13+2 theoretical parameters. This will allow for the study of 
$SU(3)$ breaking  and new-physics \mbox{effects}.

\section*{ACKNOWLEDGEMENTS}
We thank Mikolaj Misiak, and Matthias \mbox{Neubert}  
for useful  discussions and a careful reading of the manuscript.

\end{document}